\documentclass[preprint,12pt,5p,times,twocolumn]{elsarticle} 
\usepackage{epsfig}
\usepackage{color}
\journal{ }

\begin{document}
\begin{frontmatter}
\title{Grain Size and Lattice Parameter's Influence on Band Gap of SnS 
Thin Nano-crystalline Films}

\author[khalsa,sc]{Yashika Gupta}
\author[khalsa]{P.Arun \corref{cor1}\fnref{fn1}}
\ead{arunp92@physics.du.ac.in}
\author[a1]{A.A. Naudi}
\author[a1]{M.V. Walz}
\author[a1,a2]{E.A. Albanesi}

\cortext[cor1]{Corresponding author}
\fntext[fn1]{(T) +91 11 29258401 (F) +91 11 27666220}
\address[khalsa]{Department of Electronics, S.G.T.B. Khalsa College,\\ 
University of Delhi, Delhi 110007, INDIA}
\address[sc]{Department of Electronic Science, University of Delhi-South
Campus,\\ New Delhi 110021, INDIA}
\address[a1]{Facultad de Ingenieria, Universidad Nacional de Entre Rios,\\ 
3101 Oro Verde (ER), Argentina}
\address[a2]{Instituto de Fisica del Litoral (CONICET-UNL),\\ 
Guemes 3450, 3000 Santa Fe, Argentina}

\begin{abstract}
The parameters influencing the band gap of tin sulphide thin nano-crystalline 
films have been investigated. Both grain size and lattice parameters are
known to influence the band gap. The present study initially investigates
each contribution individually. The experimentally determined dependency on 
lattice parameter is verified by theoretical calculations. We also suggest
how to treat the variation of band gap as a two variable problem. The results
allow us to show dependency of effective mass (reduced) on lattice unit
volume.  
\end{abstract}


\end{frontmatter}

\section{Introduction}
The photovoltaic industry has been growing rapidly over the recent years due 
to the increasing demand for low cost and yet high efficiency solar cells. Tin 
Sulfide (SnS), a IV-VI group semiconductor having orthorhombic double
layered structure with weak Van der Waals bonds between the layers, is
considered a potential candidate due to its properties like high absorption 
coefficient (${\rm \sim 10^{4}cm^{-1}}$) and band gap (of the order of 
${\rm \sim 1.1-1.6~eV}$)~\cite{gao,yue}. SnS films are amphoteric in 
nature, i.e., they can exist either as `n'-type or `p'-type depending on the 
fabrication conditions~\cite{oga, leach}. SnS properties are also 
anisotropic~\cite{albers,gordon1,mak} which along with its amphoterism demands 
an extensive investigation into its properties.

We have noticed that the orientation with which SnS films are fabricated 
depends on the substrate it is grown on. This in turn influences the lattice 
parameters. The grain size of the films showed thickness dependence. This 
allowed us to study the properties of SnS thin films, such as band gap, as a 
function of grain size and lattice parameters. We are hence in a position to 
experimentally comment on the effective mass of the charge carriers. Finally, 
we have compared our experimental results with the theoretical calculations 
made.

\section{Experimental}
Thin SnS films of varying thicknesses were fabricated by thermal evaporation 
of SnS pellets on optically flat glass and ITO substrates (150~nm thick layer 
of Indium tin oxide grown on glass substrate) maintained at room temperature 
using a Hind High Vac (12A4D) thermal evaporation coating unit at vacuum better 
than ${\rm 4 \times 10^{-5}}$~Torr. The starting material was 99.99\% pure
SnS powder provided by Himedia (Mumbai). The thickness of the films were 
measured using Dektak surface profiler (150). The structural analysis of the 
samples were done using X-ray Diffractometer (Bruker D8 X-ray Diffractometer) 
operating at 40~KV, 40~mA with ${\rm CuK\alpha}$ radiation (λ=1.5406~\AA) 
and Transmission Electron Microscopy (Technai T30U Twin). The optical 
absorption and transmission spectra of the films were recorded using an 
UV-Vis Double Beam Spectrophotometer (Systronics 2202) over the range of 
300-1000~nm. The slow evaporation rate ensured the films were essentially
defect free and exhibited a n-type conductivity, as was confirmed by 
Hot-probe measurements.

\section{Results and Discussion}

\begin{figure}[h!!!]
\begin{center}
\epsfig{file=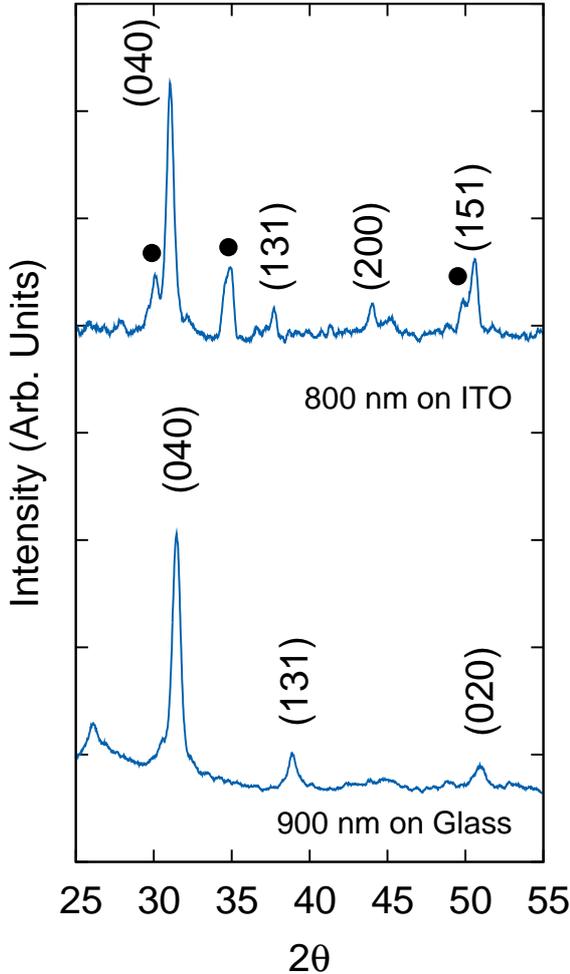, width=5.5in, angle=-90}
\end{center}
\vskip -0.6cm
\caption{X-ray diffraction pattern of SnS films of comparable thicknesses
grown on glass and ITO substrates. Filled circles indicate peaks of ITO
substrate.}
\label{fig.2}
\end{figure}

\subsection{The Structural and Morphological Analysis}
In this study, we have compared n-type SnS films grown on glass and ITO 
substrates of different thicknesses. The X-Ray Diffraction (XRD) profile 
for two comparable thicknesses are shown in fig.~\ref{fig.2}. The nature of
pattern remained the same for increasing thickness with only exception in 
case of films fabricated on ITO substrates where the ITO peaks contribution 
decreased with increasing thickness. Both
diffraction patterns matched well with the orthorhombic structure reported in 
ASTM card 83-1758 which reports the lattice parameters as a=4.148\AA,
b=11.48\AA\, and c=4.177\AA. However, the diffraction peaks corresponding to 
the (040), 
(131) and (020) planes were prominent for films grown on glass substrates 
while the films on ITO substrates shows the peaks corresponding to the (040), 
(131), (200) and (151) planes. 

\begin{figure}[h!!!]
\begin{center}
\epsfig{file=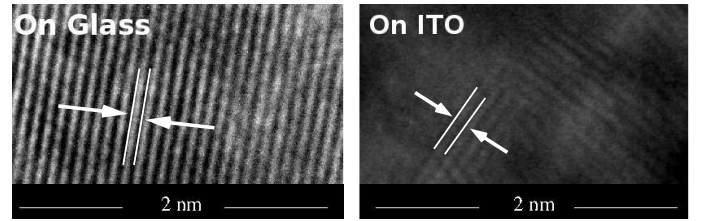, width=3.5in, angle=-0}
\end{center}
\caption{Tramsmission Electron Microscope images compare the layered
structure of SnS films on glass and ITO substrates.}
\label{fig.2tem}
\end{figure}

To investigate further, we have studied the samples using High 
Resolution Transmission Electron Microscope (HRTEM). The HRTEM images of SnS 
film on glass and ITO substrates are shown in fig.~\ref{fig.2tem}. The
layered structure of the films is evident from the parallel lines seen. The
interplanar distances can be directly measured from these images. The
interplanar distance can also be calculated from XRD data using the
formula~\cite{cullity}
\begin{equation}
\label {eq.1}
{1\over d^2} = {h^2 \over a^2}+ {k^2 \over b^2}+{l^2 \over c^2}
\end{equation}
where a, b, c are the lattice parameters and h, k, l are the Miller indices 
given in the ASTM card. The interplanar distances evaluated from HRTEM
micrographs of samples grown on glass were found to be around 0.285~nm.
Considering that ${\rm b/4 \approx 0.285~nm}$, this would imply that the lines 
seen in the micrograph are SnS layers arranged in the `ac' plane with `b' axis 
parallel to the substrate, or we may say our films have preferred
orientation with the (040) planes perpendicular to the substrate. This
orientation promises to be mechanically stable~\cite{Vidal}.  
However, inter-planar distances measured from the
micrographs of samples grown on ITO substrates are ${\rm \approx 0.31~nm}$.
This is due to the `ac' planes making an angle of ${\rm \approx 67^o}$
caused by the `b' axis making an angle of ${\rm \approx 23^o}$ with
respect to the substrate (see fig~3).

\begin{figure}[h!!!]
\begin{center}
\epsfig{file=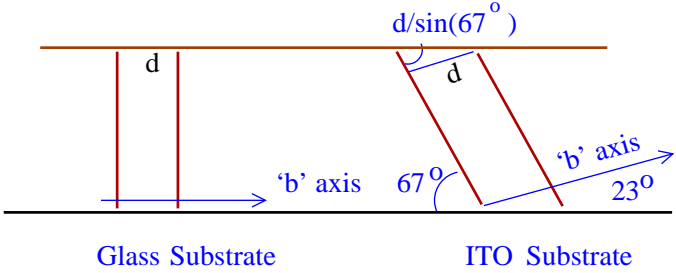, width=3.5in, angle=-0}
\end{center}
\vskip -0.4cm
\caption{Schematics shows orientation of our SnS samples on the two
substrates. }
\label{fig.33}
\end{figure}

Other then exhibiting polycrystallinity, fabricated films also exist in a 
state of stress, as is evident from the compressions or elongations 
experienced by the lattices. The lattice are said to be in a state of
stress, also refered as ``residual stress". Residual stress are remanent 
unbalanced
forces existing in the lattice due to the rapid condensation of material
during film fabrication, or due to curvatures on the substrate and/or due to
film-substrate interface etc. Residual strain can
be evaluated from the displacement of the X-ray diffraction peaks from
which lattice parameters are then evaluated. The lattice strain is given 
as~\cite{georgios}
\begin{eqnarray}
\delta={l_{OBS}-l_{ASTM} \over l_{ASTM}}
\end{eqnarray}
where `l' is the lattice parameter of the observed (subscript `${\rm OBS}$') 
and single crystal (subscript `${\rm ASTM}$') sample. Using the values of the 
elastic constants, stress can be calculated using the strain values. We have 
calculated the lattice parameters for all our samples and shall comment 
on their significance subsequently.

\begin{figure}[h!!!]
\begin{center}
\epsfig{file=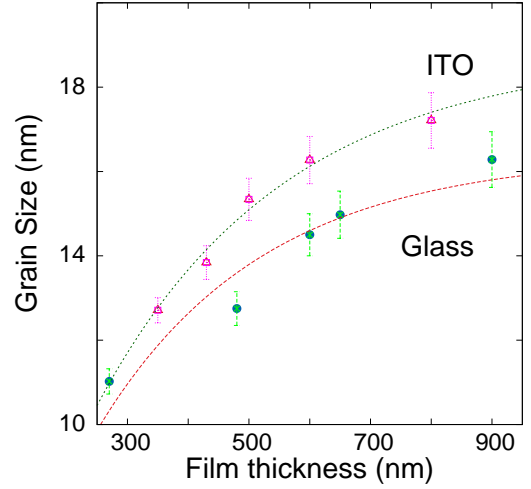, width=2.5in, angle=-90}
\end{center}
\caption{Variation in grain size with film thickness for SnS films grown on
glass and ITO substrates.}
\label{fig.1gs}
\end{figure}

To investigate the variation of grain size with substrate and film
thickness, we have calculated the average grain size of the films. The 
calculation were made using the Full Width at Half Maxima (FWHM) of the XRD 
peaks in the Scherrer’s formula 
\cite{yue,cullity,devika1}
\begin{equation}
\label{eq.2}
r={0.9 \lambda \over \beta cos \theta}
\end{equation}
where `r' is the grain size, ${\rm \beta}$ is the FWHM, ${\rm \theta}$ and 
${\rm \lambda}$ have their usual meanings. The grain sizes were found to vary 
from 11 to 18~nm depending on the film thickness in both substrates.
However, it is clear from fig.~\ref{fig.1gs} that for comparable film
thickness we get larger grains of SnS on ITO substrates. 
Devika et al~\cite{devika2} have argued that SnS nucleation is easier on ITO 
compared to glass because of its crystalline nature.

We have observed that SnS films on glass substrates have the same values for 
`b' and `c' as in single crystal. However a tensile stress exists along the
`a' direction (i.e. ${\rm a_{OBS}>a_{ASTM}}$). In contrast to this, samples
on ITO showed a compression along the `c' direction which within
experimental limits is constant for all film thicknesses. The lattice parameter
`b' remained equal to that of single crystal. Similar to the case of films
on glass, here also a tensile stress existed along the `a' axis. To 
summarize, SnS films grown on ITO have an exagerated tensile 
force acting along the `a' direction and compressive forces acting along the 
`c' direction resulting from the `b' axis not being parallel to the ITO 
substrate. This, hence, gives us an opportunity to study the properties of SnS films
and look into how variation in grain size, lattice parameter and orientation 
manifests itself on them.

\subsection{Optical Analysis}

\begin{figure}[h!!!]
\epsfig{file=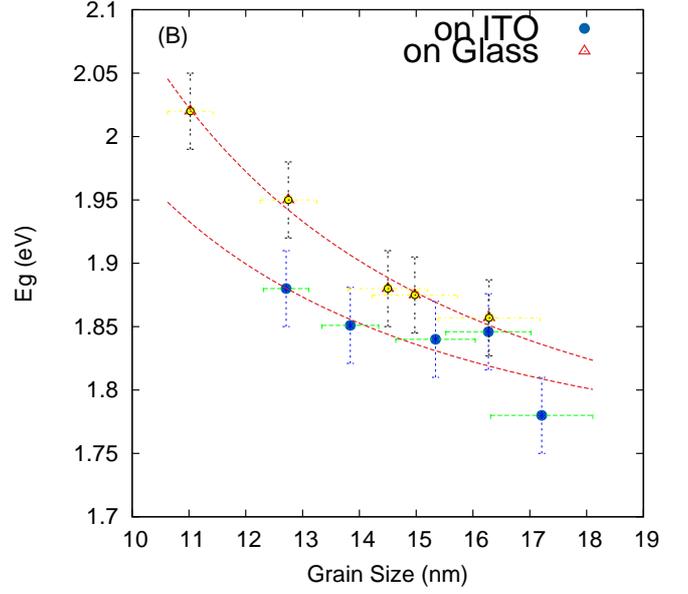, width=3.25in, angle=-90}
\caption{Variation of the band gap with grain size.}
\label{fig.4}
\end{figure}

The optical properties of a material are represented by its band gap and 
refractive index. Both these informations are evaluated from the UV-visible
absorption/ transmission spectra. The absorption coefficient 
(`${\rm \alpha}$') is calculated followed by which using the standard Tauc 
method~\cite{tauc} the band gap of films are obtained by extrapolating the 
linear part of ${\rm (\alpha h\nu)^2}$ vs ${\rm h\nu}$ plot to the `X'-axis. 
The variation of band gap with grain size of SnS films grown on ITO and 
glass substrates are shown in fig.~\ref{fig.4}. Though the trends are similar, 
the absolute values are different 
possibly due to the different crystal orientation or lattice paramter, i.e.
crystallinity, stress and orientation effect band gap
values~\cite{gordon1,ristov,pankaj}. The trend follows~\cite{brus}
\begin{equation}
\label{eq.2}
E = E_g(bulk)+{\hbar^2 \pi^2 \over 2}\left({1 \over m^*_e}+{1 \over m^*_h}\right){1 \over
r^2}
\end{equation}
or
\begin{equation}
\label{eq.2}
E = E_g(bulk)+{\hbar^2 \pi^2 \over 2\mu^*r^2}
\end{equation}
where ${\rm E_g(bulk)}$ is the band gap of SnS in bulk, `${\rm m^*_e}$' and 
`${\rm m^*_h}$' are the effective mass of electron and holes, respectively.

This result is consistent with properties induced by quantum confinement of 
charge carriers~\cite{brus}. Again this result confirms that SnS grains of
11-25~nm are in the nano-regime.
It would appear that the SnS grains with size greater than 25~nm would have
band gap similar to the bulk. Curve fits to the data points of
fig.~\ref{fig.4} give ${\rm E_g(bulk)}$ for glass and ITO as 1.707 and
1.65~eV respectively. We believe the difference in value is a result of the
different orientations in which the SnS film exists on glass and ITO
substrates. The above data also allows us to explore variation in effective
mass along different directions of SnS crystal. We however, can only comment 
on the reduced effective mass (${\rm \mu^*}$) using our experimental data.
The reduced effective mass from curve fitting is ${\rm 0.68m_o}$ and ${\rm
0.65m_o}$ for samples on glass and ITO substrates respectively, where ${\rm
m_o}$ is the rest mass of free electron.

\begin{figure}[h!!!]
\begin{center}
\epsfig{file=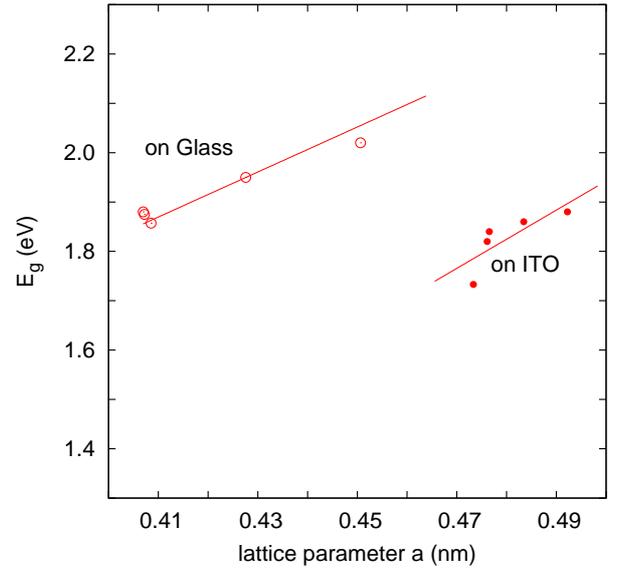, width=3.25in, angle=-90}
\end{center}
\vskip -0.4cm
\caption{A linear variation is found in SnS band gap with increasing lattice
parameter `a'.}
\label{fig.1lattice}
\end{figure}

As can be seen from Fig.~\ref{fig.1lattice}, the band gap of SnS thin films
increases with increasing lattice parameter `a'. As explained by eqn~(2), an
increase in lattice parameter is indicative of a stress acting along `a'.
Fig~\ref{fig.1lattice} shows that the stress is more in films fabricated on 
ITO substrates~\cite{ghosh}. The difference in strain (${\rm strain_{glass}<strain_{ITO}}$) 
stems from the difference in orientation between the films on different 
substrates. The data points from both the substrates do not lie on a single
line. This is indicative of the fact that SnS band gap depends on the lattice 
parameter and a second variable, possibly the grain size (from fig~5). 
Experimentally, the different
substrate used and variation in film thickness allowed us to control the two
``variables''. Theoretically, however, we can only study the variation of
band gap as a function of lattice parameters (or in other words unit cell
volume). In the next section we investigate the results of fig~6 using
theoretical calcultions.

\section{Band Structure Calculation}
\begin{figure*}[t]
\begin{center}
\epsfig{file=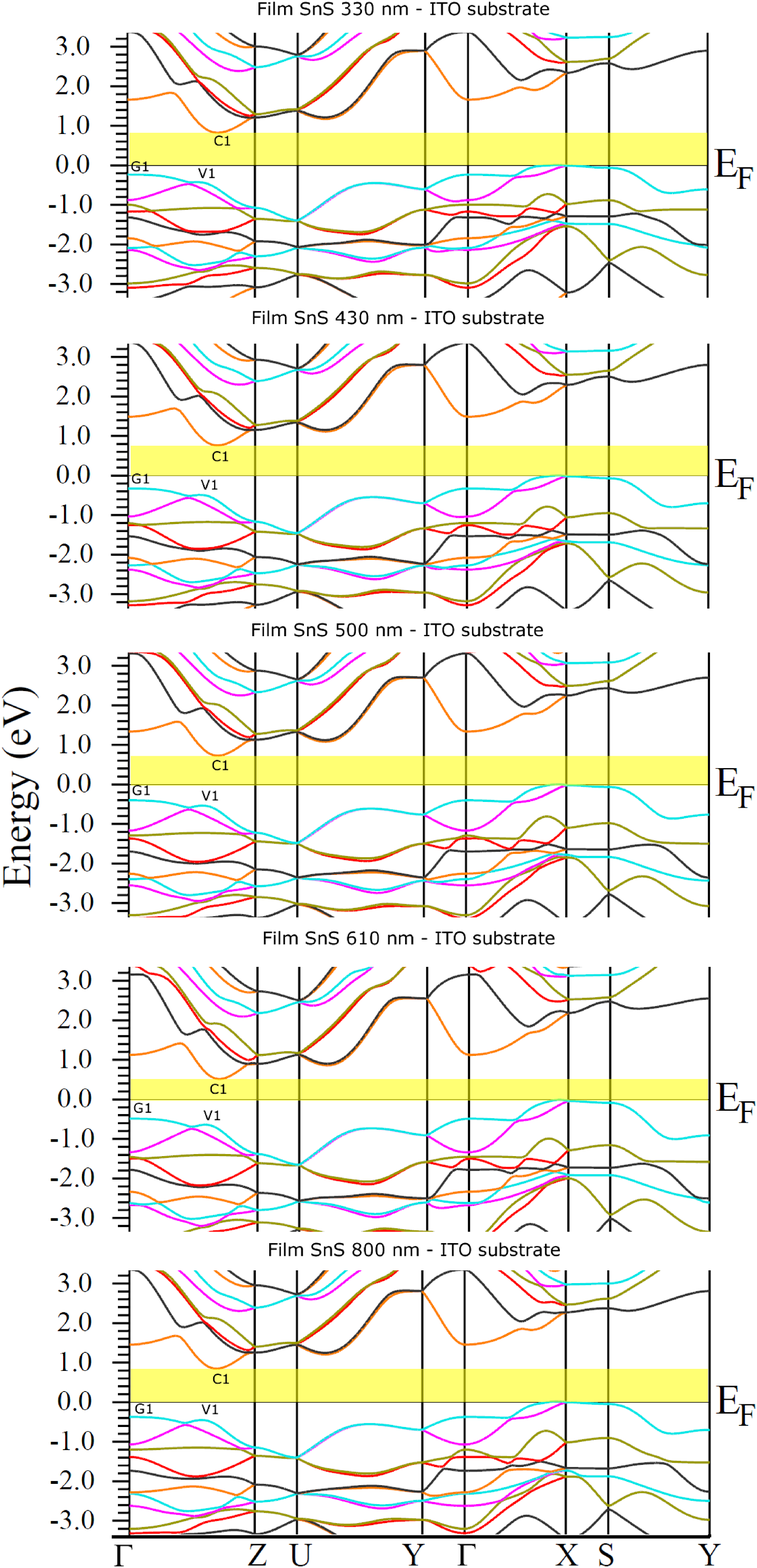, width=2.10in, angle=-0}
\hfil
\epsfig{file=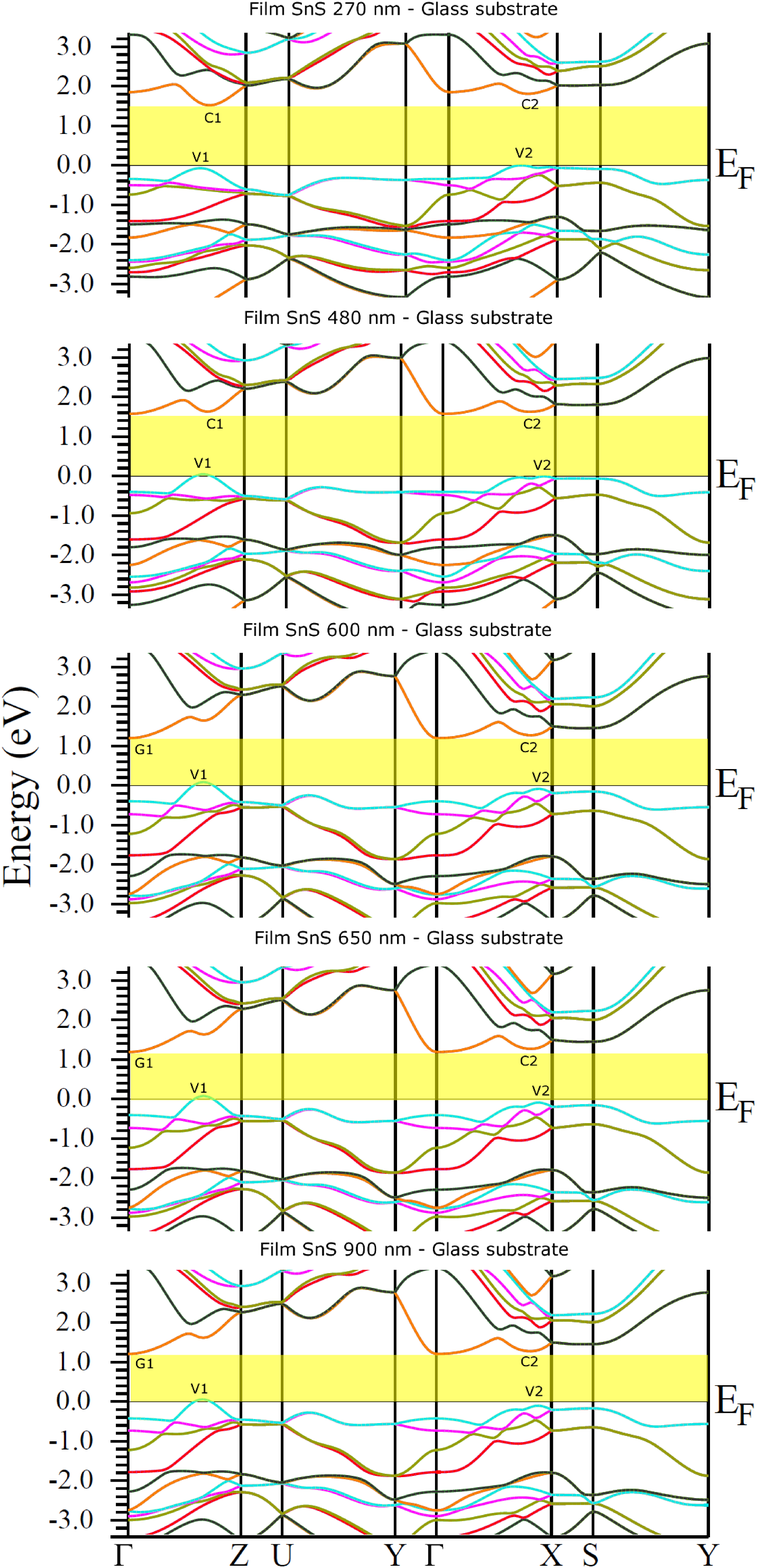, width=2.10in, angle=-0}
\caption{Results of electronic band structure calculations for SnS thin
films on ITO and glass substrates. The GW constant potential added to the
calculated band gaps are not shown in these figures. }
\end{center}
\label{fig.7}
\end{figure*}
The band structure of SnS has been theoretically evaluate
quite extensively~\cite{mak,Vidal,georgios,elm,Albanesi}. We have performed ab initio 
calculations of the SnS electronic band 
structure in the framework of the density functional theory (DFT) as 
implemented in the WIEN2k software~\cite{wien}, adopting the Engel-Vosko 
approximation for the exchange-correlation potential. We analyzed the 
energetic behavior of the compound with the inclusion of the spin orbit 
interaction. For the calculations we have used the lattice parameters 
obtained experimentally for the various SnS films with 4 Sn and 4 S atoms 
(8 atoms) per unit cell, forming two parallel zigzag chains. To achieve 
total convergence in the self-consistent calculations, we used RKMAX equal to 
7, and a converged sampling of 194 k-point in the first Brillouin Zone (BZ).

\subsection{Structure of SnS for various lattice parameters}
As stated in the preceeding section, we had deposited SnS films on ITO and 
glass substrates. We experimentally determined the lattice parameters for
these samples. Using these lattice parameters, we computed the band structure 
of SnS thin films (Fig.~7). Since the lattice 
parameter `b' showed insignificant variation with thickness in samples grown
on glass and ITO substrates, we conclude that `b' lattice parameter does not contribute 
to the variation in band gap. Also, its contribution is only along the 
${\rm \Gamma \rightarrow Y}$ direction of the BZ \cite{mak} where the ``energy 
difference'' (difference in valence band maximum and conduction band minimum) 
is too large to be considered as energy band gap associated with SnS.

However, many other valence band maximas and conduction band minimas are 
visible in the ${\rm \Gamma \rightarrow X}$ and ${\rm \Gamma \rightarrow Z}$. 
In fact, in one of the author's previous studies~\cite{Albanesi} it was 
demonstrated 
that SnS and their related IV-VI orthorhombic compounds exhibit several 
direct and indirect gaps which are close in energy and competing to form the 
band gap. This makes it difficult to precisely determine whether the band gap 
is direct or indirect (confirmed by experimental works also~\cite{44,45,46}).
Careful analysis of Fig.~7 show two comparable energy differences each 
contributing to the energy band gap of SnS films on glass and ITO substrates
respectively. The calculated band gaps are smaller than the experimental
ones due to the well known underestimation that DFT provides. This, however,
can be overcomed by including the many-body electron-electron interaction,
within the GW Hedin and Lundskvit~\cite{vv1,vv2} formalism. We have previously
calculated this correction for the semiconductor SnS~\cite{vv3}, obtaining a
constant potential value of 0.38 eV giving a  good agreement with the
experimental values. We have added this correction to all the calculated
band gaps.

In SnS films grown on ITO substrates, the smallest ``energy
difference'' is the indirect gap between the G1 maximum valence band at
${\rm \Gamma}$ point and the minimum conduction band at C1, located at 
about 1/4 from the Z point. The second energy gap is a quasi-direct energy
gap formed between the V1 relative maximum in the valence band, and the same 
minimum of the conduction band at C1. Both gaps are approximately equal and 
show the same variation with the normalised lattice's unit cell volume (fig~8). 
For obtaining the normalised lattice's unit cell volume, the stressed
lattice's unit volume is divided by the volume listed in the given ASTM Card
(henceforth lattice's unit cell volume would imply normalised lattice's unit 
cell volume).

\begin{figure}[h!!!]
\epsfig{file=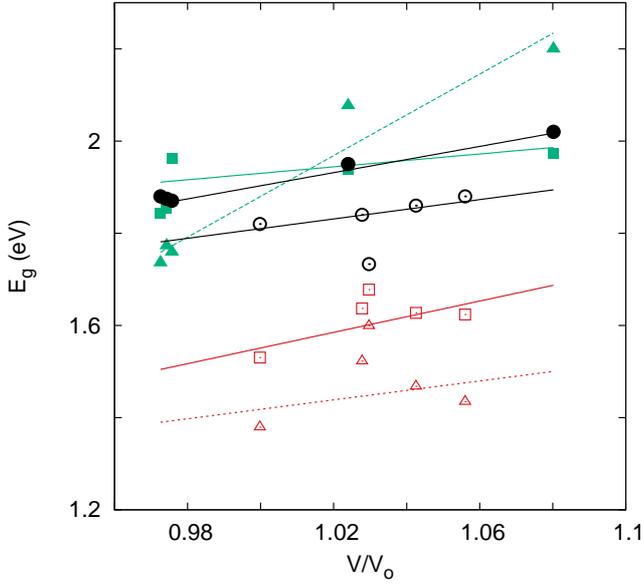, width=3.25in, angle=-90}
\caption{ Plots show the band gaps of SnS films as a function of the lattice's
unit cell volume. Full squares (trend marked by a green solid line)
represents the quasi-direct band gaps in the `c' axis direction (${\rm
\Gamma \rightarrow Z}$ direction in the BZ) while the full triangles (trend marked
by green dashed lines) gives the direct band gaps in the `a' axis 
direction (${\rm \Gamma \rightarrow X}$ direction in the BZ) for SnS films on
glass substrates. The unfilled squares (trend marked by a red solid line) 
represents the 
quasi-direct band gaps in the `c' axis direction (${\rm \Gamma \rightarrow Z}$ 
direction in the BZ) while the unfilled triangles (trend marked
by red dashed lines) gives the indirect band gaps in the `c' axis 
direction (${\rm \Gamma \rightarrow Z}$ direction in the BZ) for SnS films on
ITO substrates. For comparing, full circles and unfilled circles with black
solid lines are given which represent the experimental band gaps of SnS films
on glass and ITO substrates respectively.
}
\label{fig.44}
\end{figure}

The films on glass substrates also show two important energy gaps, they are 
quasi-direct, one along  along the ${\rm \Gamma \rightarrow Z}$ direction of
the BZ that occurs at 1/4 from Z point and second along the 
${\rm \Gamma \rightarrow X}$ direction of the BZ. The variation of
band gap with lattice's unit cell volume is shown in fig~(\ref{fig.44}). In 
fig~(\ref{fig.44}), we have included the experimental data also (black solid
and dashed trendlines are for SnS films on glass and ITO substrates
respectively). Considering that the experimental method adopted to evaluate
band gaps can not resolve the ${\rm \Gamma \rightarrow X}$ and 
${\rm \Gamma \rightarrow Z}$ contributions, the average of the band gaps 
calculated along the ${\rm \Gamma \rightarrow X}$ and 
${\rm \Gamma \rightarrow Z}$ directions also compared with those of the
experimental data. The trendlines (slope) matched well (not shown in
fig~\ref{fig.44} for brevity).

Considering that the variation in lattice parameters would result in residual 
stress effects that would manifest as pressure, we expect changes in 
conduction band, valence band shapes and band gaps~\cite{georgios,Albanesi}.
Georgies et al~\cite{georgios}, Makinistian et al~\cite{Albanesi} and Parenteau 
et al~\cite{parenteau} works show that the band gap of SnS is directly
proportional to the unit cell's volume, with band gap approaching small
values as volume decreases. Fig~\ref{fig.44} show the results of our calculations for
samples grown on ITO 
and glass respectively for different thicknesses. The linearity is in
confirmation of results in literature and experimental results highlighted
in fig~(\ref{fig.1lattice}) and fig~(\ref{fig.44}).
The theoretical work hence matches and substantiates the experimental results.
However, the theoretical calculations of band structure does not take into
account grain boundary and hence is not in a position to comment on the
variation of band gap with grain size. As stated in our section on
experimental results, the band gap depends on both the grain size and
lattice's unit cell volume (via lattice parameter). To this extent, we
believe that the two variables are acting independent of each other and are
hence separable. Eqn~(5) should then be written as
\begin{eqnarray}
E(r,V) &=& \left[A+{B \over r^2}\right]\left(m{V \over V_o}\right)\nonumber\\
&=& \left({V \over V_o}\right)E_g(bulk)+{\hbar^2\pi^2V \over
2\mu^*V_o}\left({1 \over r^2}\right)\label{hyp}
\end{eqnarray}

\begin{figure}[h!!!]
\epsfig{file=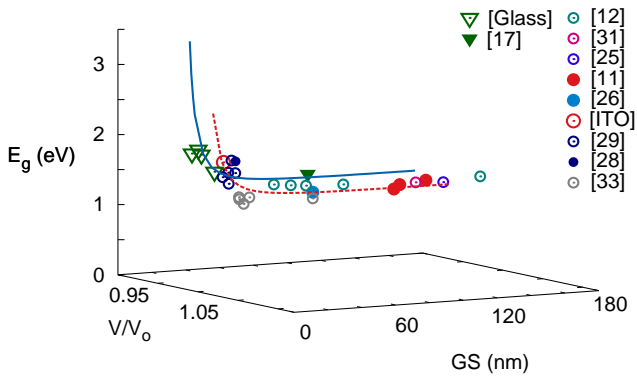, width=2.15in, angle=-90}
\caption{The band gap is plot as a function of normalized lattice volume and
grain size (GS). The points arrange in two curves, namely 
${\rm V/V_o \approx 0.98}$
and ${\rm V/V_o \approx 1}$ (data points indicated with triangles and
circles respectively). The source of references are indicated. (Color is
available online)
}
\label{fig.3dfinal}
\end{figure}

Fig~(\ref{fig.3dfinal}) shows a three dimensional plot of our experimental
data along with data collected from various literature
\cite{devika1,devika2},\cite{devika3}-\cite{add2} on SnS thin
films. It is seen that the data points arrange themselves along two family
of graphs, namely on ${\rm V/V_o}$~${\rm \approx}$ 0.975 and 0.995 for various 
grain sizes. A large volume of data exists for ${\rm V/V_o \approx 0.995-1}$. 
Again if we consider eqn~(5) to be valid, then a family of curves should not 
have existed. If forced to curve fit, then identical values of
${\rm E_g(bulk)}$ and ${\rm {\hbar^2\pi^2 \over 2\mu}}$ should returnned. 
This is not the
case. A look into the derivation of Brus~\cite{brus} eqn~(5) would show that 
it considers the electron's Coulombic interactions only. This interaction 
become prominant only when the electrons are confined 
in grains whose dimension are on the nano-scale. In large crystals, these
interactions are neglected since the electrons are far apart and their 
interactions
are only with the lattice potential whose periodicity is related to the
lattice parameter. The curvature of the band structures thus formed due
to the electron lattice potential interaction, is related to the magnitude of
the effective mass. Our results suggest, that even at the nano-scale, the
electron lattice potential interaction can not be neglected and has to be
considered along with the electron-electron interaction potential. Results
emerging from fig~(\ref{fig.3dfinal}) suggest that resulting Schrodinger
equation is solvable by separable variable method leading to an expression
like eqn~(6).

Correcting for the lattice parameter, we have reduced effective
mass for (${\rm V/V_o} \approx $) 0.975 and 0.995 as ${\rm 0.247m_o}$ and
${\rm 0.279m_o}$, respectively. This is more near to the values given in
literature by theoretical calculations~\cite{albers,Vidal,georgios} and 
confirmed by our own theoretical calculations. The data points on the two set 
of curves of 
fig~(\ref{fig.3dfinal}) are from various sources and include those on glass
and ITO substrates. Hence, the trend now is fully explained by grain size
and lattice parameter. 

\section{Conclusions}
The variation in experimental data of SnS thin nano-crystalline film's band gap 
in the present work and of those from the literature are completely explained 
by considering it to be a two variable problem. The band gap depends on the
grain size and the lattice volume through the lattice parameters. Theoretical
calculations have been used where ever possible to corroborate the results.
The results are of significance in Material Science where material
characteristics are manipulated as per requirement.

\section*{Acknowledgement}
Discussions with Sukanta Dutta, Department of Physics, S.G.T.B. Khalsa
College is gratefully acknowledged. One of the authors (YG) would like to
acknowledge DST (India) for the fellowship (Fellowship No. IF131164) awarded 
under its Inspire Scheme. The authors (AAN, MVW and EAA), acknowledge the 
financial support from the Universidad Nacional de Entre Ríos (UNER) and the 
Consejo Nacional de Investigaciones Científicas y Tecnicas (CONICET), 
Argentina.

\end{document}